# A Robust Asynchronous Newton Method for Massive Scale Computing Systems

Travis Desell

Department of Computer Science
University of North Dakota
Grand Forks, ND 58202, USA
travis.desell@gmail.com

Malik Magdon-Ismail, Heidi Newberg, Lee A. Newberg, Boleslaw K. Szymanski, Carlos A. Varela

Network Science and Technology Center
Rensselaer Polytechnic Institute
Troy, NY 12180, USA

*Abstract*—Volunteer computing grids offer supercomputing levels of computing power at the relatively low cost of operating a server. In previous work, the authors have shown that it is possible to take traditionally iterative evolutionary algorithms and execute them on volunteer computing grids by performing them asynchronously. The asynchronous implementations dramatically increase scalability and decrease the time taken to converge to a solution. Iterative and asynchronous optimization algorithms implemented using MPI on clusters and supercomputers, and BOINC on volunteer computing grids have been packaged together in a framework for generic distributed optimization (FGDO). This paper presents a new extension to FGDO for an asynchronous Newton method (ANM) for local optimization. ANM is resilient to heterogeneous, faulty and unreliable computing nodes and is extremely scalable. Preliminary results show that it can converge to a local optimum significantly faster than conjugate gradient descent does.

## I. Introduction

Volunteer computing grids can offer significant levels of computing power at very low costs. As an added benefit, many volunteers continually upgrade their hardware, so computing power of a volunteer computing project increases over time, while at best a supercomputer stays the same. However, utilizing these extremely large scale systems involves significant challenges in overcoming heterogeneous, faulty and even malicious hosts. The computations performed are also usually limited embarrassingly parallel bag-of-tasks type work. In many cases, effectively utilizing a volunteer computing system requires rethinking the algorithms involved.

In previous work, the authors have shown that asynchronous versions of evolutionary algorithms can be effectively run on volunteer computing systems, such as MilkyWay@Home [1]. While evolutionary algorithms can effectively find global (or near global) solutions to difficult computational problems with many local optima, they are not nearly as efficient as local optimization methods in more well behaved search spaces with a single optimum. Additionally, after finding the general area of the global optimum, they may then take a very long time to converge to the solution.

This work explores an asynchronous version of the Newton method, which has traditionally been avoided in smaller scale computing systems. By using regression to calculate the search direction and then using a randomized line search, it is possible to perform an efficient local optimization method on a large scale computing system. The asynchronous Newton method (ANM) presented is extremely scalable and tolerant to heterogeneous, faulty hosts. As part of FGDO, it also uses BOINC [2] to validate results from volunteers, providing protection from malicious hosts. Preliminary results show that it converges to a solution in significantly less iterations than conjugate gradient descent and can scale to a massive scale computing system like MilkyWay@Home which currently consists of around 35,000 volunteered hosts (see http://boincstats.com).

## II. Iterative Local Optimization

In traditional local optimization scenarios, conjugate gradient descent (CGD) [3] or quasi-Newton (QN) methods [4] are typically favored over a standard Newton method, as they require fewer function evaluations to converge to the local optimum. Both types of methods start at a point, $\vec{x}$, in the parameter space. For each iteration, both the CGD and QN methods will then calculate a precise approximation of the gradient, $\nabla$, of the function $f$ at point $\vec{x}$. The $i^{th}$ value of the gradient vector, $\nabla f(\vec{x})_i$ is:

$$\nabla f(\vec{x})_i = \frac{f(\vec{x}+\vec{s}_i^{\,0}) - f(\vec{x}-\vec{s}_i^{\,0})}{2s_i} \quad (1)$$

where $\vec{s}_i^{\,0}$ is a vector of all zeros, except with a user defined step size $s_i > 0$ as the $i^{th}$ element. For example, given a uniform step vector of length $n+1, \vec{s}_{i=0..n}^{\,0} = 0.1, \vec{s}_2^{\,0}$ would be *[0, 0, 0.1, 0, .., 0]*. CGD will use this gradient to update a stored conjugate gradient, while QN methods use the gradient to refine their approximation of a Hessian matrix (the second-order partial derivatives of a function).

Following the calculation of the gradient, a direction, $\vec{d}$, for a line search is chosen (starting at $\vec{x}$). CGD uses the conjugate gradient as the direction, while QN methods use the inverse of the approximate Hessian multiplied by the gradient. It is also possible to perform an inexact line search using Wolfe conditions [5], which compute an acceptable step length by which to multiply the direction to choose the next point, $\vec{x}_{next}$. Following the exact or inexact line search, the next iteration will begin using it as the initial point.

In a standard Newton method, a precise approximation of the Hessian matrix is calculated in a similar manner to the gradient, also using the user defined step vector, $\vec{s}$. The value of the Hessian matrix at row $i$ and column $j$ of function $f$ at point $\vec{x}$ is:

$$H(f(\vec{x}))_{i,j} = \frac{1}{4 s_i s_j} \left[ f(\vec{x} + \vec{s}_i^0 + \vec{s}_j^0) - f(\vec{x} - \vec{s}_i^0 + \vec{s}_j^0) \right.$$
$$\left. - f(\vec{x} + \vec{s}_i^0 - \vec{s}_j^0) + f(\vec{x} - \vec{s}_i^0 - \vec{s}_j^0) \right] \quad (2)$$

Then the direction, $\vec{d}$, for the line search is calculated as:

$$\vec{d} = -H(f(\vec{x}))^{-1} \nabla f(\vec{x}) \quad (3)$$

As with CGD and QN methods, the line search is used to calculate $\vec{x}_{next}$ and the process repeats until the optimization process cannot progress any further. From (1), the calculation of a gradient for CGD or a QN method only requires $2n$ function evaluations, where $n$ is the number of parameters. However, from (2) a numerical calculation of the Hessian requires $4n^2$ function evaluations, which can be slightly reduced to $4n^2-n$ by avoiding a function evaluation on the diagonal. Numerically calculating the Hessian provides the most accurate approximation for calculating the line search direction, which results in less overall iterations of the optimization method. However, given the drastic increase in function calculations required to numerically calculate the Hessian, it is easy to see why the standard Newton method is usually avoided for optimization.

### III. An Asynchronous Newton Method

It is possible to calculate the Hessian as well as the gradient using regression instead of numerically. In a large scale computing system, especially in a heterogeneous and faulty computing system such as BOINC, using regression provides many advantages over a numerical calculation. In the latter, if any one of the processes fails in calculating its function evaluation, the entire computation stalls until that function evaluation is calculated by a new node. Further, the numerical calculation only scales to $4n^2-n$ concurrent function evaluations, and CGD and QN methods have even less scalability, only $2n$ function evaluations. Using regression, it is possible to obtain a more accurate Hessian by using additional points in its calculation, which can be especially beneficial if the user defined step vector, $\vec{s}$, is not well specified.

First, function evaluations must be performed for a random set of $m$ points, $x^{i=0..m-1}$. These are calculated around the initial point, $\vec{x}'$, using the user defined step vector, $\vec{x}' \pm \vec{s}$. These points are then used to create a vector of function evaluation results, $\vec{y}_i = f(\vec{x})^i$, as well as an initial matrix $X$ using the different points used in calculating $\vec{y}$. With $n$ denoting the number of parameters in the search space, row $i$ and column $j$ of $X$ are:

$$X_{i,0} = 1$$
$$X_{i,1..n} = \vec{x}_j^i, \quad j = 0..n-1$$
$$X_{i,n+1..2n} = \frac{1}{2} \vec{x}_j^i \vec{x}_j^i \quad j = 0..n-1$$
$$X_{i,2n+1..n^2+n} = \frac{1}{2} \vec{x}_j^i \vec{x}_k^i \quad j = 0..n-1, k = j+1..n-1$$

A vector containing the gradient and the Hessian, $\vec{B}$ can be calculated as follows:

$$\vec{B} = (X^T X)^{-1} X^T y \quad (4)$$

Then the gradient and Hessian are:

$$\nabla_i = \vec{B}_{i+1}, i = 0..n-1$$
$$H_{i,i} = \vec{B}_{n+i+1}, i = 0..n-1 \quad (5)$$
$$H_{i,j} = \vec{B}_{2n+1+ni+j}, i = 0..n-1, j = i+1..n$$
$$H_{i,j} = H_{j,i}$$

In (4), the matrix $X$ must be at least a square matrix, so the number of function evaluations required to do the regression calculation is at least $n^2+n$. Because these points are all randomly generated, it is possible on BOINC to send out a large number of random points and then gather results of the function evaluations as they are reported. When enough results have been gathered, the regression can be performed and the gradient and Hessian from (4) and (5) can be used to calculate the direction for a line search using (3).

### IV. An Asynchronous Line Search

Calculating the gradient and Hessian with the above methods provides varying amounts of scalability, with the asynchronous method providing the most of it. However, most line search methods, such as a logarithmic line search or Brent's method [6], are inherently sequential, performing one function evaluation at a time because future function evaluations require the fitness evaluation of the previous function evaluation(s). In the case of MilkyWay@Home, a single function evaluation can take a day or longer, having any client or the server perform the line search would severely degrade the performance of the optimization method. Even worse, on volunteer computing systems there is no guarantee that a client will ever return a result, so having this kind of dependency is infeasible.

A randomized line search can be used to determine the area where the next set of random points for the gradient and Hessian calculation is selected. A randomized line search is extremely simple, selecting random points, $\vec{x}_{random}$, within the bounds of the search space and a user specified optimum bound, $\alpha_{min}$, and maximum bound, $\alpha_{max}$. These bounds are also limited in each iteration by the user specified borders of the search space, $b_{min}$ and $b_{max}$. They are increased or decreased so no point along the directional line could be outside the search space. These random points are selected along the direction calculated by (3), where $r$ is a random number between zero (inclusive) and one (non-inclusive), and $\vec{x}'$ is the center of the previous regression:

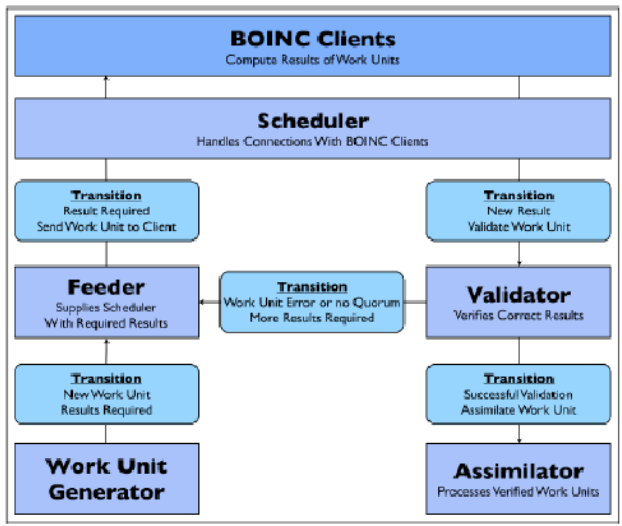

Fig. 1. The BOINC architecture consists of multiple server-side daemons which handle work generation, validation and scheduling of tasks, or *workunits*, to clients.

$$\vec{x}_{random} = \vec{x}' + \alpha_{min}\vec{d} + r(\alpha_{maz} - \alpha_{min})\vec{d} \qquad (6)$$

This randomized line search can also be performed asynchronously. As clients request work, more random points can be generated without any dependencies on other points previously sent out for the line search. When a sufficient number of results have been reported, we can select the best point and use that as the center of the next Hessian and gradient calculation.

While this approach is quite simple, this method has some significant benefits in that there are no explicit dependencies that would make heterogeneous function evaluation time or the failure of any client problematic. It is also very scalable, increasing the number of random points increases the accuracy of the line search. Further, it is possible to escape from local optima using this randomized approach (as shown in Section VI), which is not possible using traditional iterative line search methods, which only find the nearest optimum.

## V. IMPLEMENTAION

ANM and the randomized line search were implemented in FGDO for use on BOINC. FGDO provides a specialized combination of BOINC's assimilator, validator and work generator daemons, because in asynchronous optimization methods, the workunits generated depend on up-to-date results that have been received from clients which are processed by the assimilator. Further, in asynchronous optimization the number of workunits that require validation can be even further reduced by only validating results that will be used to generate new workunits [7].

Using FGDO, ANM works as follows. First, the user selects the initial central point $\vec{x}$ and step vector $\vec{s}$. Workunits are generated around the current point and sent out to volunteers. When a user specified number of function evaluations were calculated and validated by the BOINC clients, they are used to perform the regression. Then for the randomized line search, workunits are generated on a line along the direction calculated by the regression and sent out to volunteers. After a user specified number of function evaluations are calculated and validated, the result with the best fitness is chosen as the center for the next regression. The regression and line search process repeats until the search stops making acceptable progress.

## VI. RESULTS

Preliminary results for ANM have been gathered using MilkyWay@Home. They were evaluated using data from the Sloan Digital Sky Survey [8]. A model for one tidal stream of the Sagittarius dwarf galaxy disruption within those stripes and the background distribution of stars in the Milky Way Galaxy were fit to the observed data. The model has 8 different parameters used in the optimization and the data sets consisted of 92,000 to 112,000 stars. For each iteration of ANM, 1000 points each were used to calculate the regression and to perform the randomized line search. As volunteers requested work, more points for evaluation would be sent out until 1000 results had been received.

Figure 2 shows the progress of two ANM searches for stripe 79 and 86, started from randomly selected points close to the global optima. Stripe 79 converges to optima quite quickly, in 5 iterations. Stripe 82 takes longer, converging within 20 iterations. This corresponds to 10,000 and 40,000 function evaluations, respectively. Optimization of stripes 79 and 86 have been studied significantly in previous work [9,10,11] and starting from similar positions take hundreds of iterations to converge to the global optimum with similar accuracy using conjugate gradient descent.

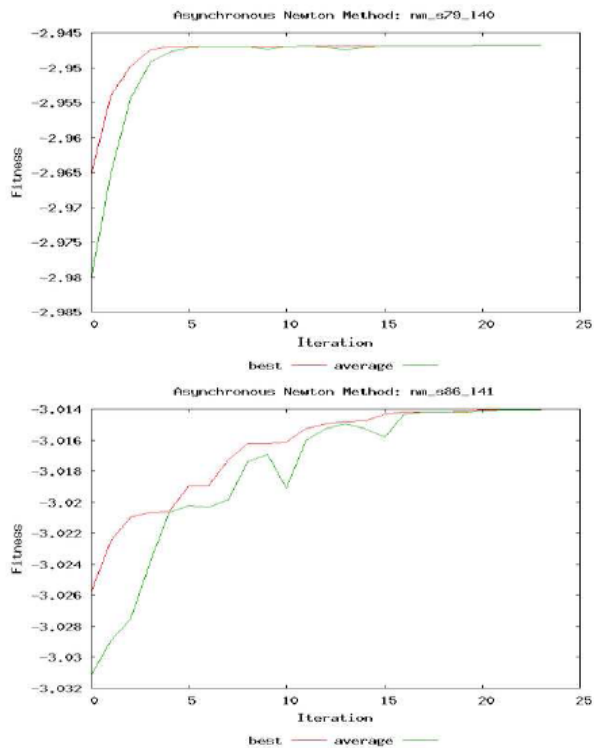

Fig 2. The progress of ANM performing parameter optimization on two datasets from the SDSS, showing the best and average fitness from the randomized line search of each iteration.

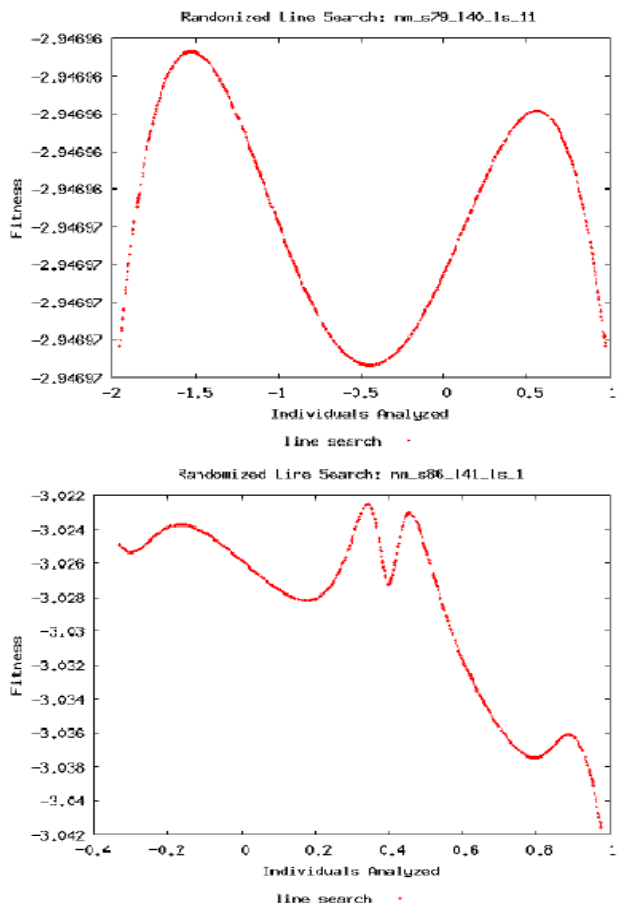

Fig. 3. Examples of the randomized line search breaking out of local optima.

While this method may take fewer function evaluations to find a solution than ANM, it has significantly less parallelism; in this case 16 function evaluations could be done concurrently to calculate the gradient, while the line search has no parallelism at all. Because of this, ANM can have significantly less time to solution given enough volunteers.

A further benefit from this approach is that the randomized line search has the potential to break out of local optima. Figure 3 demonstrates this with a figure of the points evaluated during line searches for stripe 79 and 82. They demonstrate that while the regression did pick a direction which corresponds to the overall gradient of the line search, using a traditional line search would not pick the overall best point, as they would start at 0 and only move to the nearest best point. This has the potential to even further decrease the number of iterations the Newton method takes to reach the nearest optimum.

## VII. CONCLUSIONS

Traditionally, using the standard Newton method in optimization has been avoided due to the excessive number of function evaluations it requires to calculate a search direction. However an asynchronous variation of the Newton method using regression can take advantage of the massive amounts of parallelism provided by volunteer computing systems. This asynchronous implementation is extremely scalable and resilient to faulty and unreliable nodes. It also has the additional advantage of being able to break out of local optima using a randomized line search. Preliminary results show that this method has the potential to converge to a local optimum in significantly fewer iterations than conjugate gradient descent, which can drastically reduce the time to solution when using a large enough computing system.

This work opens up many avenues for future research. It may be possible to further reduce the time to solution by using Wolfe conditions to do an inexact line search [5] or by using the error values from the regression to further refine the range of the randomized line search. It may also be possible to reduce the time to find a global optimum by initially using various asynchronous evolutionary algorithms [10] to come close to the global optimum and then utilize ANM to refine the accuracy of the final value. Further analysis of ANM may even show that it can compare to asynchronous evolutionary algorithms in finding a global optimum due to the randomized line search being able to escape local optima. As such, we feel that this is a promising method warranting deeper study.

ACKNOWLEDGMENT

The authors would like to thank all the volunteers at MilkyWay@Home and DNA@Home who have made this work possible. This work has been partially supported by the NSF under Grants No. 0612213, 0947637, and 10-09670.

REFERENCES

[1] T. Desell, D. Anderson, M. Magdon-Ismail, B. S. Heidi Newberg, B. Szymanski and C. Varela, "An analysis of massively distributed evolutionary algorithms," in The 2010 IEEE congress on evolutionary computation (IEEE CEC 2010), Barcelona, Spain, July 2010.

[2] P. Anderson, E. Korpela, and R. Walton, "High-performance task distribution for volunteer computing." in e-Science. IEEE Computer Society, 2005, pp. 196–203.

[3] M. R. Hestenes and E. Stiefel, "Methods of conjugate gradients for solving linear systems," Journal of Research of the National Bureau of Standards, vol. 49, no. 6, December 1952.

[4] J. Bonnans, J. Gilbert, C. Lemarechal, and C. A. Sagastizabal, Numerical optimization, theoretical and numerical aspects. Springer, 2006.

[5] J. Nocedal and S. Wright, Numerical Optimization. New York, NY: Springer Verlag, 1999.

[6] R. P. Brent, Algorithms for Minimization without Derivatives. Englewood Cliffs, NJ: Prentice-Hall, 1973.

[7] T. Desell, M. Magdon-Ismail, B. Szymanski, C. Varela, H. Newberg, and D. Anderson, "Validating evolutionary algorithms on volunteer computing grids," in The 10th IFIP international conference on distributed applications and interoperable systems (DAIS). Amsterdam, Netherlands: Springer-Verlag, June 2010.

[8] J. Adelman-McCarthy, "The 6th Sloan Digital Sky Survey Data Release, http://www.sdss.org/dr6/," July 2007, apJS, arXiv/0707.3413.

[9] [9] T. Desell, N. Cole, M. Magdon-Ismail, H. Newberg, B. Szymanski, and C. Varela, "Distributed and generic maximum likelihood evaluation," in 3rd IEEE International Conference on e-Science and Grid Computing (eScience2007), Bangalore, India, December 2007, pp. 337–344.

[10] T. Desell, "Asynchronous global optimization for massive scale computing," Ph.D. dissertation, Rensselaer Polytechnic Institute, 2009.

[11] N. Cole, "Maximum likelihood fitting of tidal streams with application to the sagittarius dwarf tidal tails," Ph.D. dissertation, Rensselaer Polytechnic Institute, 2009.